\documentclass[aip,cha,reprint,floatfix]{revtex4-1}
\usepackage[dvipdfm]{graphicx}
\usepackage{amsmath,amssymb,amsfonts}
\usepackage{latexsym}
\usepackage{verbatim}

\newcommand{\length}{{\rm length}\ }

\begin{document}
\title{On repellers in quasi-periodically forced logistic map system}
\date{3 March 2014}

\author{Tsuyoshi Chawanya}
\email{chawanya@ist.osaka-u.ac.jp}
\author{Takafumi Sakai}
\affiliation{Graduate School of Information Science \& Technology, Osaka University, Toyonaka, JAPAN.}

\begin{abstract}
We propose a method to identify and to locate "repellers'' in
quasi-periodically forced logistic map (QPLM), using a kind of Morse
decomposition of nested attracting invariant sets.  In order to obtain
the invariant sets, we use an auxiliary 1+2-dimensional skew-product
map system describing the evolution of a line segment in the phase
space of QPLM.  With this method, detailed structure of repellers can
be visualized, and the emergence of a repeller in QPLM can be detected
as an easily observable bifurcation in the auxiliary system. In
addition to the method to detect the repellers, we propose a new
numerical method for distinguishing a strange non-chaotic attractor
(SNA) from a smooth torus attractor, using a correspondence between
SNAs in QPLM and attractors with riddled basin in the auxiliary
system.
\end{abstract}
\maketitle

\begin{quotation}
The topological structure of invariant sets as well as its variation
(bifurcation) is one of the most basic and essential issues for
understanding the behavior of a dynamical system.  However, it is
sometimes not easy to obtain invariant sets in non-autonomous systems,
and that might make our understanding of those systems rather
limited. In this paper we propose a method to identify and to locate
repellers in quasi-periodically forced logistic map system, and
exhibit the obtained images of repellers and bifurcation diagrams.  It
would be helpful for developing better understanding of mechanisms for
the emergence of complex behaviors in quasi-periodically forced
systems from dynamical system point of view.
\end{quotation}

\section{introduction}

Quasi-periodically forced systems could be considered as one of the
simplest class of ``essentially non-autonomous'' dynamical systems.
In this class of systems, surprisingly complex dynamical phenomena,
including rather common existence of strange non-chaotic
attractors(SNA), were reported in early 1980s \cite{Grebogi-etal-84,
  Kaneko-84a, Kaneko-84b}. Since then the feature and the mechanism of
such phenomena have been extensively studied by many researchers with
various background, and related phenomena have been uncovered in
relatively wide variety of systems
\cite{Feudel-etal-06,Ding-etal-89,Prasad-etal-98,Prasad-etal-07,Senthilkumar-etal-08,Mitsui-etal-12,Suresh-etal-13}.
Major part of the theoretical researches on the phenomena have been
done mainly for the cases with systems with ``forced'' invariant set
induced by something like symmetry
\cite{Keller-96,Yalcinkaya-Lai-97,Yang-01,Glendinning-02a,Glendinning-02b,Alseda-Misiurewics-08,Bjerklov-12,Bjerklov-09}.
On the other hand, also for the systems without such forced invariant
sets, rigorous results are obtained however mainly in reversible
1+1-dimensional systems: the existence of SNA is indicated for the
case where the existence of continuous attractor is excluded by
topological constraint\cite{Hunt-Ott-01,Kim-ETAL-03} and the emergence
of SNA at non-uniform saddle-node bifurcation is proven
\cite{Jaeger-09}.  As for higher dimensional and/or irreversible
systems, also where large part of the numerical studies have been
done, there still many phenomena which have not been understood
clearly.

When a quasi-periodic external force is applied to such systems that
possibly exhibit chaotic behavior even without external forcing, it
seems natural to expect that unstable invariant sets with complicated
structure would play essential role in the
dynamics\cite{Badard-08,Glendinning-ETAL-06}.  In quasi periodically
forced circle map system with relatively strong modulation (thus it is
no longer reversible), existence of such complicated structure is
indicated using densely distributed winding number of the
orbits\cite{Glendinning-ETAL-09}.  Also in numerical researches,
repelling invariant sets with complicated structure are observed, for
example, as a basin boundaries between two distinct attractors, and
the boundary crisis has been identified as one of the routes to SNA.
Thus such repelling invariant sets are considered to be directly
relevant with the emergence of SNA at least for such cases.

In some other cases, the existence of repelling sets is also strongly
suggested from observation of the parameter dependence of the size or
the smoothness of the attractor (i.e., crisis-like drastic enlargement
of an attractor or ``fractalization'' of a torus), while there seem to
be no visible basin boundary in its neighborhood.  In such cases it is
not trivial task to identify and to locate such presumed repellers.
One possible approach to locate such invariant sets is to consider a
perturbation to the invariant sets in the system without external
forcing.  However, this method is applicable only for limited class of
invariant sets (i.e., a torus or composition of multiple tori).
Another useful approach is to approximate the quasi-periodic forcing
with periodic one\cite{Pikovsky-Feudel-95,Chastell-etal-95}, and thus
approximate 1+1-dimensional skew-product system with 1-parameter
family of 1-dimensional autonomous systems. This method, known as
rational approximation (RA) method, have been used as a major tool to
observe such invariant sets in numerical researches.  It has been used
for the classification of the bifurcation to create SNA in QPLM
\cite{Heagy-Hammel-94,Nishikawa-Kaneko-96,Witt-etal-97,Prasad-etal-97,Feudel-etal-98,Kim-etal-03},
and corresponding types of bifurcation are observed also in QP-forced
higher dimensional systems
\cite{Osigna-Feudel-00,Kim-Lim-05,Lim-Kim-05}.  Though the observed
features of the ``repeller'' obtained in RA are consistent with the
qualitative changes of the behavior of the attractors on the whole, RA
cannot reproduce the topological structure of the genuine ``repeller''
verbatim, and the correspondence of the bifurcations in QP-forced
systems and those obtained by RA is somewhat obscure.

In this report, we propose a method to locate and to obtain the images
of repellers in QP-forced logistic map system(QPLM) including those
with non-trivial structures, and exhibit some of the numerical results
obtained with this method.  Our results are consistent with the
bifurcation scenario that have been inferred from RA method or from
direct observation of the attractors, and provide some novel
perspective to bifurcation phenomena where various repellers are
involved.

Our method is presented in section 2.  We introduce an auxiliary
dynamical system describing the evolution of a segment, that is given
as a 1+2 dimensional skew-product map system, and describe outline of
the method to identify ``repeller'' of the QPLM.  Some details of the
algorithm is given in the the appendix.  Some of the numerical results
obtained with this method are demonstrated in section 3, including
detailed phase diagram exhibiting a complicated bifurcation structure
near TDT(torus doubling termination) critical point, as well as images
of repellers with non-trivial structure.  Summary and short discussion
will be given in the last section.

\section{Method}

In this section, we will explain our method to obtain the images of
``repellers'' in QPLM.  In the first place, we introduce concrete form
of the system and auxiliary ``segment map'', a 1+2- dimensional
skew-product map that describes the evolution of a ``segment'' in the
phase space of QPLM.  Then we describe the method to obtain repellers
in QPLM.  Some technical details of the algorithm to obtain the images
of repellers, together with a method to distinguish SNA from
``smooth'' torus will be given in appendix.

\subsection{QPLM and ``segment map''}

Here we consider the quasi-periodically forced logistic (quadratic)
map in the following form,
$M|\Omega\rightarrow\Omega$,$(\Omega:=T^1([0,1))\times{\mathbf R})$,
  $M(\theta_n,x_n)=(\theta_{n+1},x_{n+1})$
\begin{eqnarray}
\theta_{n+1} &=& (\theta_n + \omega) \ \mbox{mod}\ 1,\cr
x_{n+1} &=& a - x_n^2 + \epsilon \cos(2\pi\theta_n).
\label{EQ:QPLM}
\end{eqnarray}
$a$ and $\epsilon$ are treated as control parameters, and
$\omega$ is fixed as $(\sqrt{5}-1)/2$ in this paper.

In order to examine the stability of a point in $\Omega$, it would be
a natural approach to consider the temporal evolution of its
neighborhood.  Regarding the trivial neutral stability in the $\theta$
direction, it seems reasonable to think about the evolution of a
vertical segment that contains the target point. Fortunately the
evolution of a vertical segment can be written concretely as a
(1+2)-dimensional piece-wise polynomial skew-product map system.

Thus here we introduce an auxiliary ``segment map'' system, given as
$\tilde M\ |\ \tilde \Omega\rightarrow\tilde \Omega,\ \tilde
\Omega=T^1([0,1))\times({\mathbf R}\times{\mathbf R}_+), {\mathbf
    R}_+=\{x\in{\mathbf R}|x\geq 0\}$, $\tilde
  M(\theta_n,z_n,w_{n})=(\theta_{n+1},z_{n+1},w_{n+1})$
\begin{eqnarray}
\theta_{n+1} &=& (\theta_n + \omega) \ \mbox{mod}\ 1,\cr
z_{n+1} &=& \left\{
\begin{array}{ll}
a - z_n^2 -w_n^2 + \epsilon \cos(2\pi\theta_n),& (|z_n|\geq w_n)\cr
a - (|z_n|+w_n)^2/2 + \epsilon \cos(2\pi\theta_n),& (|z_n|<w_n)
\end{array}
\right.\cr
w_{n+1} &=&  \left\{
\begin{array}{ll}
2|z_n|w_n,& \ (|z_n|\geq w_n)\cr
(|z_n|+w_n)^2/2,& \ (|z_n|<w_n)
\end{array}
\right.
\end{eqnarray}
Here $\theta$ is the ``phase'' of the external force as in the
QPLM(\ref{EQ:QPLM}). $z$ and $w$ correspond to the center position and
the half-length of the segment respectively.  Although the derivative
of this map has discontinuity on $z=0$, we have observed no symptom of
pathological phenomena due to this discontinuity.

Note that, the subspace of zero-length segments
($\tilde\Omega_0=\{w=0\}\subset\tilde\Omega$) is kept invariant by the
map $\tilde M$, and the restriction of $\tilde M$ on $\tilde\Omega_0$
could be naturally identified with QPLM ($M$ on $\Omega$).

We sometimes call a subset of the phase space with common $\theta$
value as a ``fiber'' (both for QPLM and for the segment map
system). For a point $u=(\theta,z,w)\in\tilde\Omega$, the length of
the corresponding segment in $\Omega$ ($= 2w$) is written as
$\length(u)$.  For a point $u=(\theta,z,w)$ in $\tilde \Omega$, we let
$\underline u$ to represent a set (segment) $\{(\theta,x)\in\Omega
\ |\ z-w\leq x\leq z+w\}$.  Similarly, for $U$ which is a subset of
$\tilde\Omega$, let $\underline U$ stands for $\bigcup_{u\in
  U}\underline u \subset \Omega$.  We call $\underline U$ as the
``shadow'' of $U$. For an invariant set (that may or may not be an
attractor) $X\in W$ where $W$ may be either $\Omega$ or
$\tilde\Omega$, we let $B(X)$ denote the basin of attraction $\{u\in W
\ |\ \omega(u)\subset X\}$, where $\omega(u)$ denotes the
$\omega$-limit set of the orbit starting from $u$.

\subsection{A method to locate repellers}

Now let us describe the outline of the method to detect and to locate
``repellers''.  Our basic strategy is to find a set of points that are
transversely unstable and non-wandering, using the map describing
temporal evolution of a line segment.

As for the transverse stability of a point, we observe asymptotic
behavior of an orbit (of the segment map) starting from a segment
(with positive length) containing the point, and check if the length
of the segment would converge to 0 or not.  If it converges to 0, that
indicates no transversely unstable point is contained in the initial
segment.  On the other hand, if the length of the orbit starting from
the segment has a positive limsup value, transversely unstable
point(s) presumably exist in the initial segment.

The asymptotic behavior of the orbit of the segment map gives some
information about the recurrence property of the points in the initial
segment as well.  As the $\omega$-limit set of an orbit of the segment
map is an invariant set, its shadow determines a corresponding
invariant set of the original QPLM.  If the initial segment (i.e., a
segment in QPLM corresponding to the shadow of the initial point of
the segment map) is a subset of this invariant set, there should exist
non-wandering points in the initial segment.

Thus we are interested in such point ($p$) that satisfies
\begin{itemize}
\item $\omega$-limit set of any orbits (of $\tilde M$) starting from a
  point corresponding to non-zero length segment containing the point
  $p$ is detached from $w=0$ plane,
\item the shadow of the $\omega$-limit set contains the point $p$
\end{itemize}
Above mentioned $\omega$-limit set for starting point corresponding to
sufficiently short segments is expected to be one of the attractors of
the segment map. Let $A$ denote the attractor. The ensemble of orbits
starting from a neighborhood of the original point $p$ should be dense
in the shadow of the attractor $A$.  Thus the shadow of the attractor
$A$ could be regarded as a ``basin of repulsion'' of the repelling
non-wandering point $p$.

There can be multiple repelling non-wandering points with the common
``basin of repulsion''.  The set of those repelling non-wandering
points consist a transitive set.  Thus the set as a whole could be
identified as a ``repeller''.  We let $R(A)$ denotes a repeller with
``basin of repulsion'' $A$.

Basically what we try to obtain is a image of the section of repellers
on a target fiber.  In order to do this, we try to carry out the
following steps.
\begin{itemize}
\item Obtain the list of pull back attractors of the segment map on
  the target fiber,
\item Find pull back attractors that correspond to a segment
whose proper subset corresponds to another pull back attractor,
\item Look for such points which are in the shadow of the outer
  attractor and whose corresponding $\omega$-limit set does not belong
  to any of inner attractor's shadow.
\end{itemize}
The obtained set of points would be the section of the repeller whose
basin of repulsion is given by the shadow of the outer attractor.  In
practice, we try to obtain the image of multiple sections of the
repellers using the information of pull back attractors on one fiber.
The detail of the procedure is described in the appendix.

It should be noted that a pull back attractor on a fiber does not
simply correspond to a section of the attractor in autonomous
2-dimensional system.\cite{Kloeden-Rasmussen-11} If $A\subset \Omega$
is an asymptotically stable attractor of $M$ in the view the 2-dim
autonomous system, boundary of $A$ would consist of graph(s) of
continuous and piece-wise smooth function of $\theta$. In such cases,
the intersection of $A$ with a certain fiber $\{\theta=c\}$ should
coincide with a pull back attracting set on the fiber (which may
consists of multiple pull back attractors).  On the other hand, for
the cases with SNA, although pull-back attractor(s) for almost all
fibers can be obtained just like the above case, the shape of their
union over the fibers is not continuous.  Thus SNA does not simply
correspond to an asymptotically stable attractor of the 2-dimensional
autonomous systems.

It is not easy to distinguish these two types of pull back attractors
from simple observation.  For the case of SNA, however, there should
exist a repelling orbit (that should be a part of a repeller) in
arbitrarily small neighborhood of the pull back
attractor\cite{Sturman-Stark-00}.  Thus we could expect that there
exist trajectories of $\tilde M$ starting from a point in an
arbitrarily small neighborhood of an ``SNA'' but finally attracted to
another attractor (of $\tilde M$) that corresponds to the basin of
repulsion of the repeller which is located in touch with the
attractor(SNA).  Thus, it is expected that SNAs exhibit a kind of
sensitivity of final state against arbitrarily small perturbation in
the neighborhood of the attractor in $\tilde M$, like those attractors
with ``riddled'' basin\cite{Alexander-ETAL-92,Yang-01}.  We try to
detect this presumed type of sensitivity of final state in $\tilde M$
expected for SNAs, by checking whether a small perturbation for $w$
component of $\tilde M$ would make the destination of the trajectory
to another attractor (in $w>0$ region) or not.  Thus we introduce a
``perturbed'' version of segment map $\tilde M^+$, given as $\tilde
M^+ = \tilde W^+ \circ \tilde M$ where $\tilde W^+ \ | \ \tilde \Omega
\rightarrow \tilde \Omega$, $\tilde
W^+(\theta,z,w)=(\theta,z,\min\{w,\delta^+\})$, where $\delta^+$
denotes the amplitude of the perturbation.  ( $\delta$ is set as
$5.0\times 10^{-7}$ in the examples shown in this report.)

\section{Numerical results}

In this section, we present some of the numerical results, namely,
phase diagrams in $(a,\epsilon)$-plane and visualized images of
attractors and repellers for some selected parameter values.

\begin{figure}
\includegraphics[clip,width=0.45\textwidth]{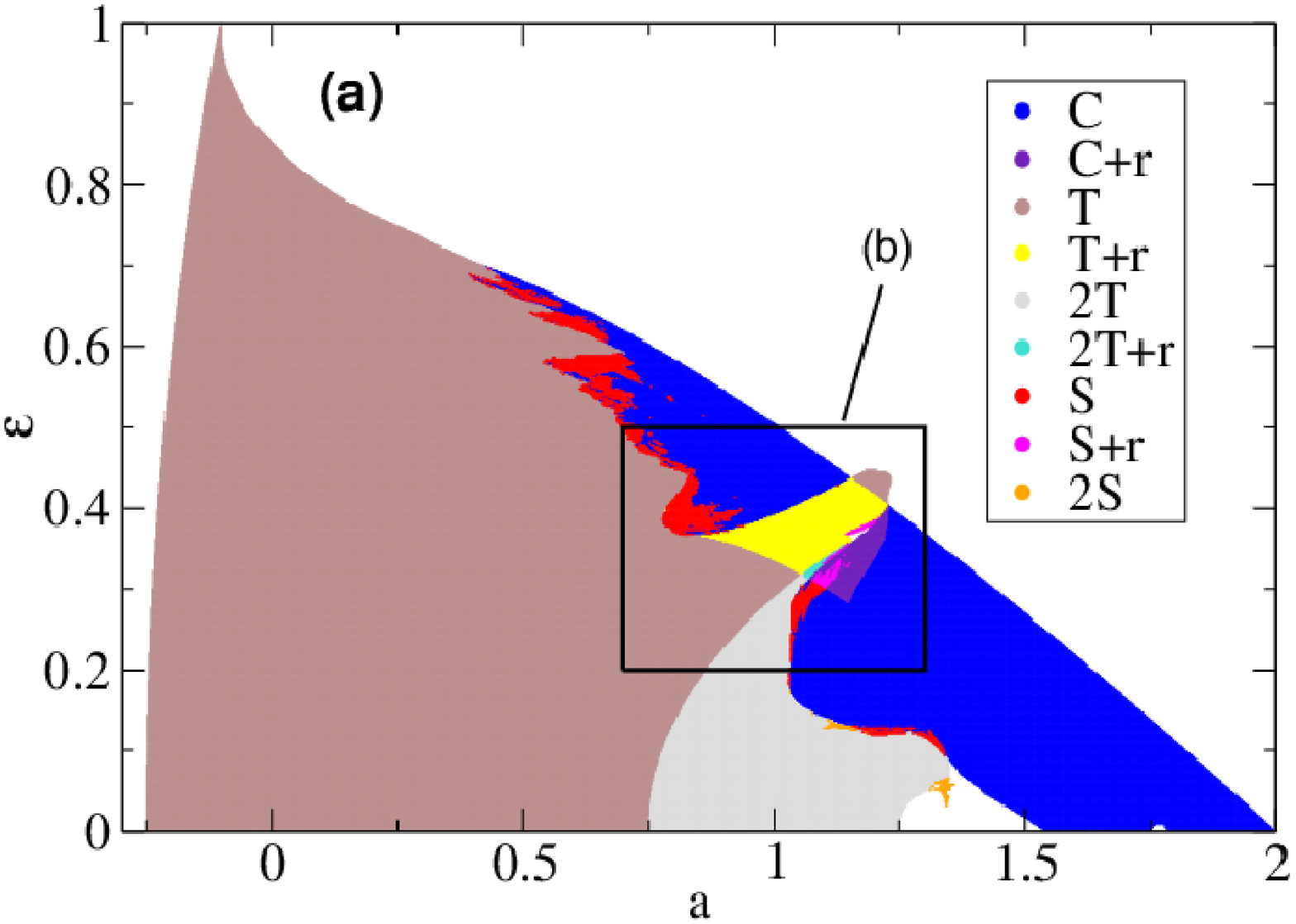}
\includegraphics[clip,width=0.45\textwidth]{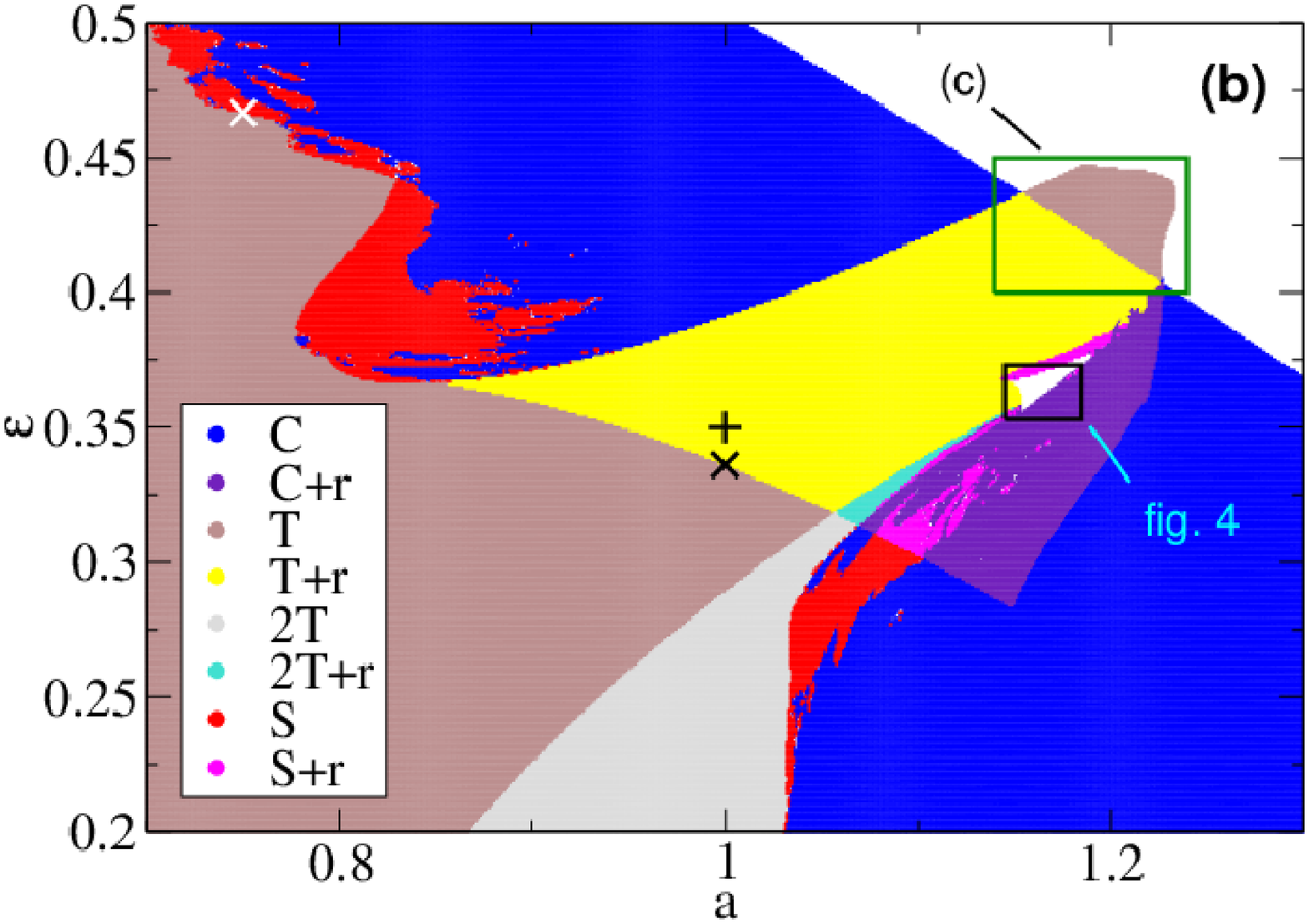}
\includegraphics[clip,width=0.45\textwidth]{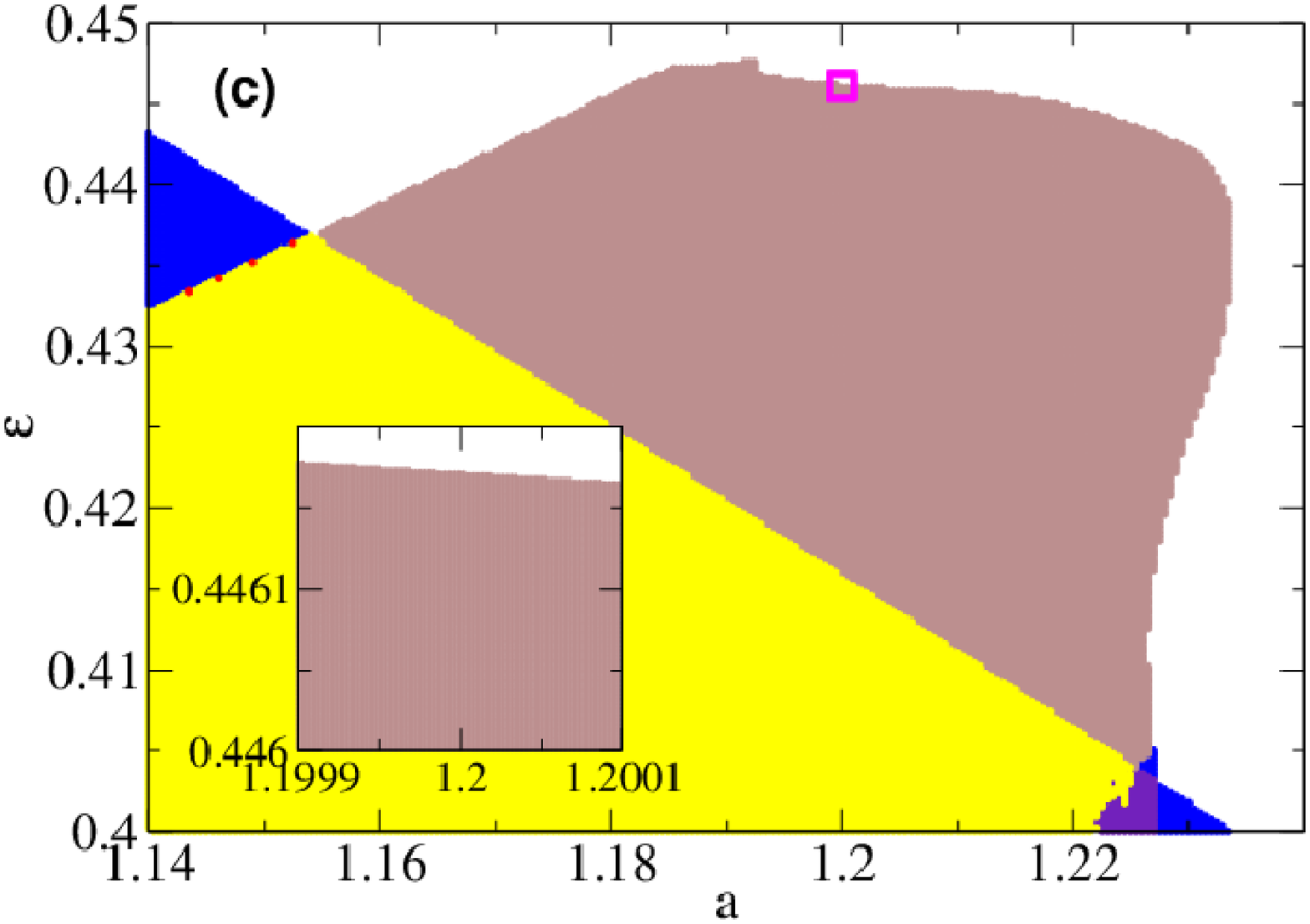}
\caption{Phase diagram in $a$-$\epsilon$ plane} Marks in the box (b)
indicate parameter values corresponding to the following
figures. Black star for fig. \ref{FIG:typical-example}, black $\times$
for fig. \ref{FIG:repeller-onset}, white $\times$ for
fig. \ref{FIG:fractalized}. Those plots are obtained with mesh size
$(\Delta a,\Delta\epsilon)$ taken as $(0.002,0.002)$ for (a),
$(0.001,0.001)$ for (b), $(0.0005,0.0002)$ for (c) and
$(0.000002,0.000001)$ for the inset of (c).  The observed type of the
attractor or the repeller is indicated with color as illustrated in
the legend box in box (b).  C, T, 2T and S respectively stand for
chaotic, torus, period 2 tori and strange non-chaotic attractor, and
``+r'' indicates the existence of a ``non-trivial'' repeller which is
not a part of the basin boundary with other attractors.
\label{FIG:DIAGRAM}
\end{figure}

The obtained phase diagram (Fig.\ref{FIG:DIAGRAM}) seems similar to
those obtained in the preceding studies. However, since our criterion
to identify an attractor as SNA is different from those used in the
preceding studies (based on the divergence of the estimated phase
sensitivity), the position of the obtained borderline between smooth
attracting torus and SNA has some discrepancy.  It would be notable
that we observed no SNA near the border of the boundary crisis which
is associated with the disappearance of bounded attractor in the
bump-like region near $(a,e)=(1.2, 0.42)$ (as is shown in box (c)),
which is consistent with the natural expectation that SNA appears only
AT the boundary crisis for such cases.

\begin{figure}
\includegraphics[clip,width=0.48\textwidth]{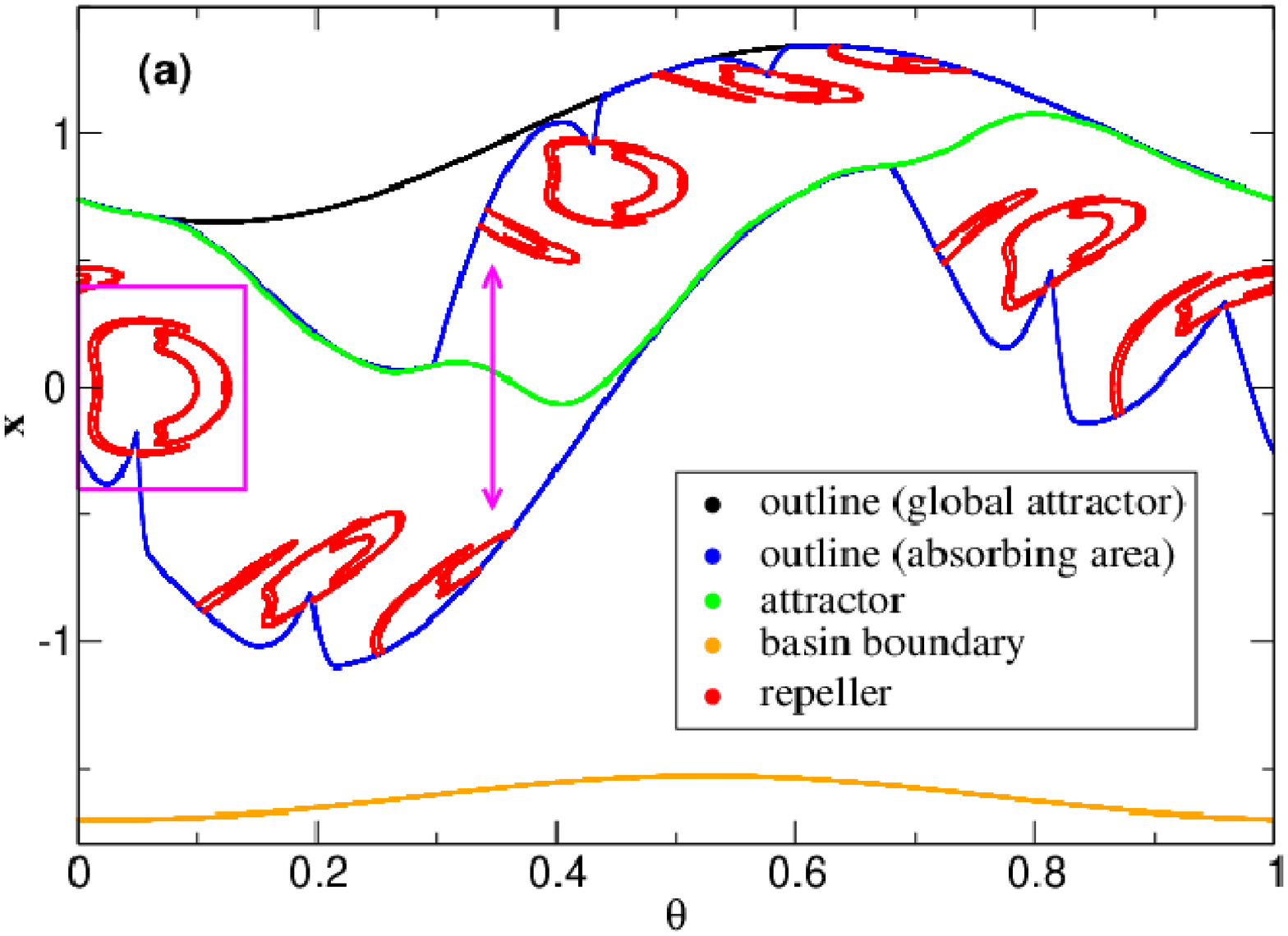}\hfill
\includegraphics[clip,width=0.48\textwidth]{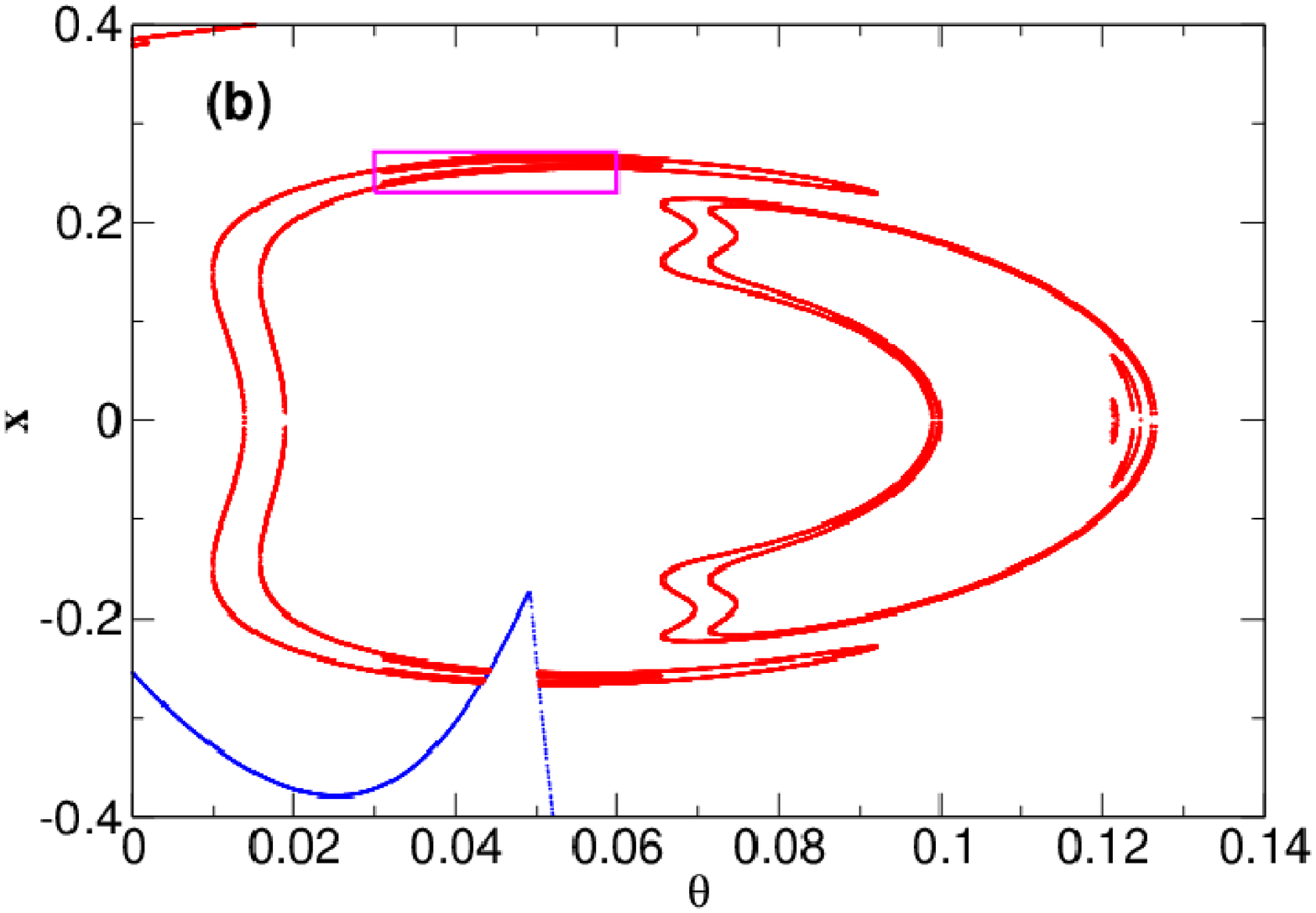}\hfill
\includegraphics[clip,width=0.48\textwidth]{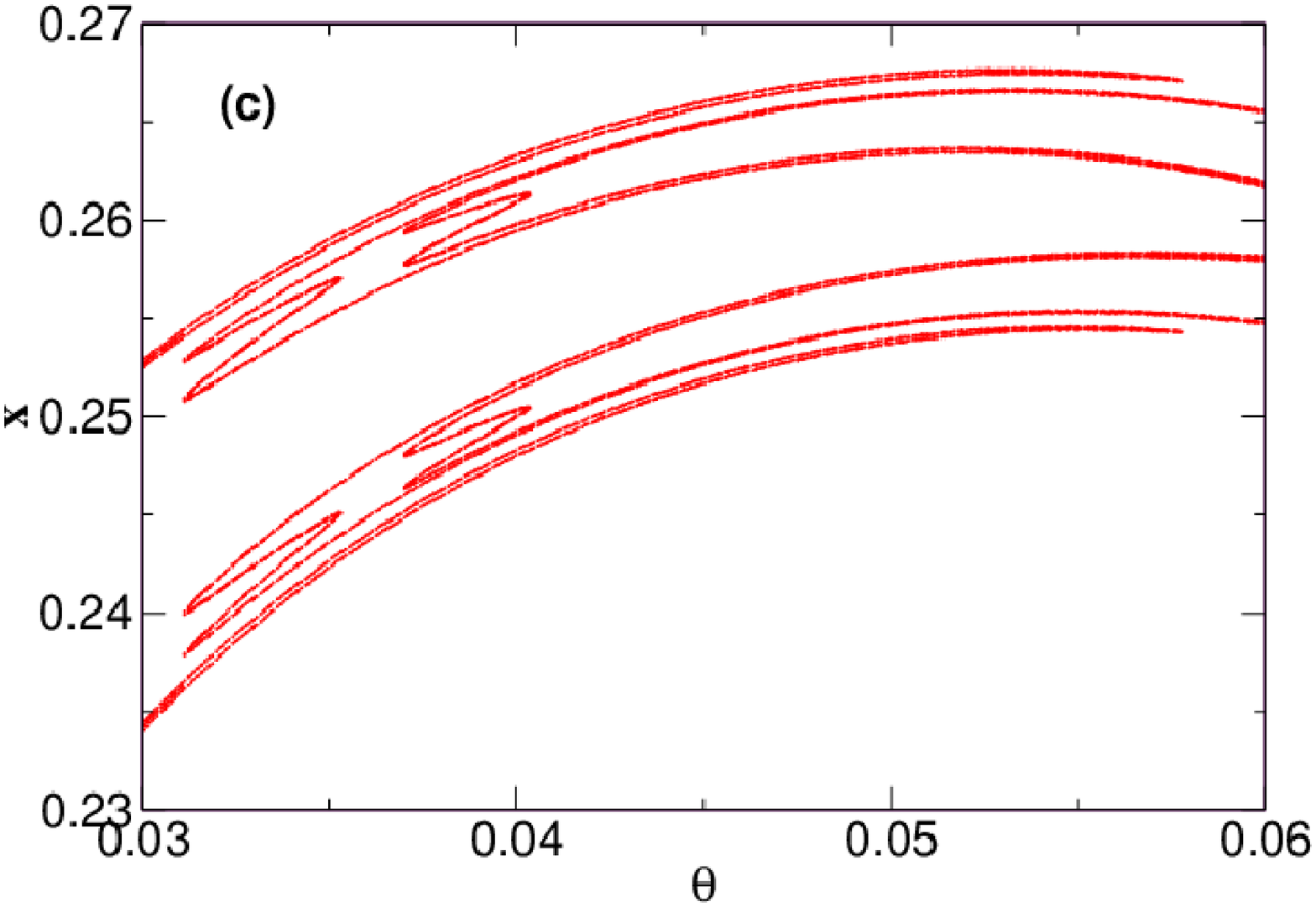}\hfill
\caption{An example of repeller with non-trivial structure observed at
  $(a,\epsilon)=(1.0,0.35)$.  Thin apertures in the rim of ring like
  structures are visible in enlarged view in boxes (b) and (c). Data
  of pull-back attractors and repellers on 10000 target fibers in
  $\theta\in[0,1]$ are used. \label{FIG:typical-example}}
\end{figure}

An example of the repellers with non-trivial structure (observed as
``ring shaped'' repellers in RA method) is exhibited in
Fig. \ref{FIG:typical-example}.  Besides the apparently fractal
layered structure, thin apertures can be observed in the rim of the
rings (in boxes b and c).  These thin apertures are mapped to the
large gap near $\theta=0.35$ (indicated by a two-way arrow in box a),
after 15 (for box b) and 28 (for box c) iterations.

The above result suggests that the repeller has a family of infinitely
many apertures (with progressively thin widths), and thus when
adequately coarse grained it could appear to consist of finite number
of chunks.  Thus the approximation for the repeller obtained by RA
could be a good approximation for the repeller in quasi-periodically
forced system, in spite of the fact that the set of finite number of
rings cannot be invariant in the quasi-periodically forced system.

In the next figure (Fig. \ref{FIG:repeller-onset}), we chose the
parameter near the emergence of the ring-like shaped repeller, and
visualized the shape of the repeller and the attracting invariant set
corresponding to the basin of repulsion of the repeller.  Sequence of
inward spikes are observed in the boundary curve of the basin.  We
think that these spikes are a symptom of a non-uniform saddle-node
bifurcation, where the attractor of $\tilde M$ (which corresponds to
the basin of the ring-like shaped repeller) collide with a saddle.
Thus it is surmised that, at the bifurcation of the disappearance of
the repeller, the boundary would touch the last component of the
repeller on infinitely many fibers simultaneously, while on typical
fibers the repeller of QPLM with fractal-like structure is still
retained.  Thus it is conjectured that the repeller would have
strictly positive lyapunov exponent even at the onset.

\begin{figure}
\includegraphics[clip,width=0.48\textwidth]{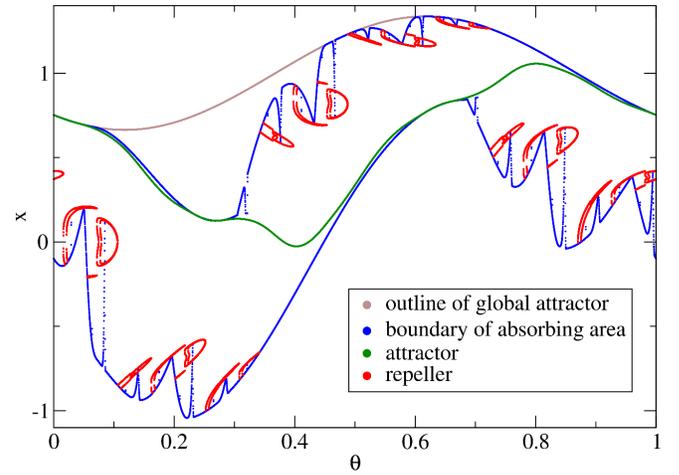}
\caption{Repeller and its basin of repulsion, near the emergence of
  the repeller $(a,\epsilon)=(1.0,0.336057731)$. Data of 10000 target
  fibers are plotted together. \label{FIG:repeller-onset}}
\end{figure}

\begin{figure}
\includegraphics[clip,width=0.48\textwidth]{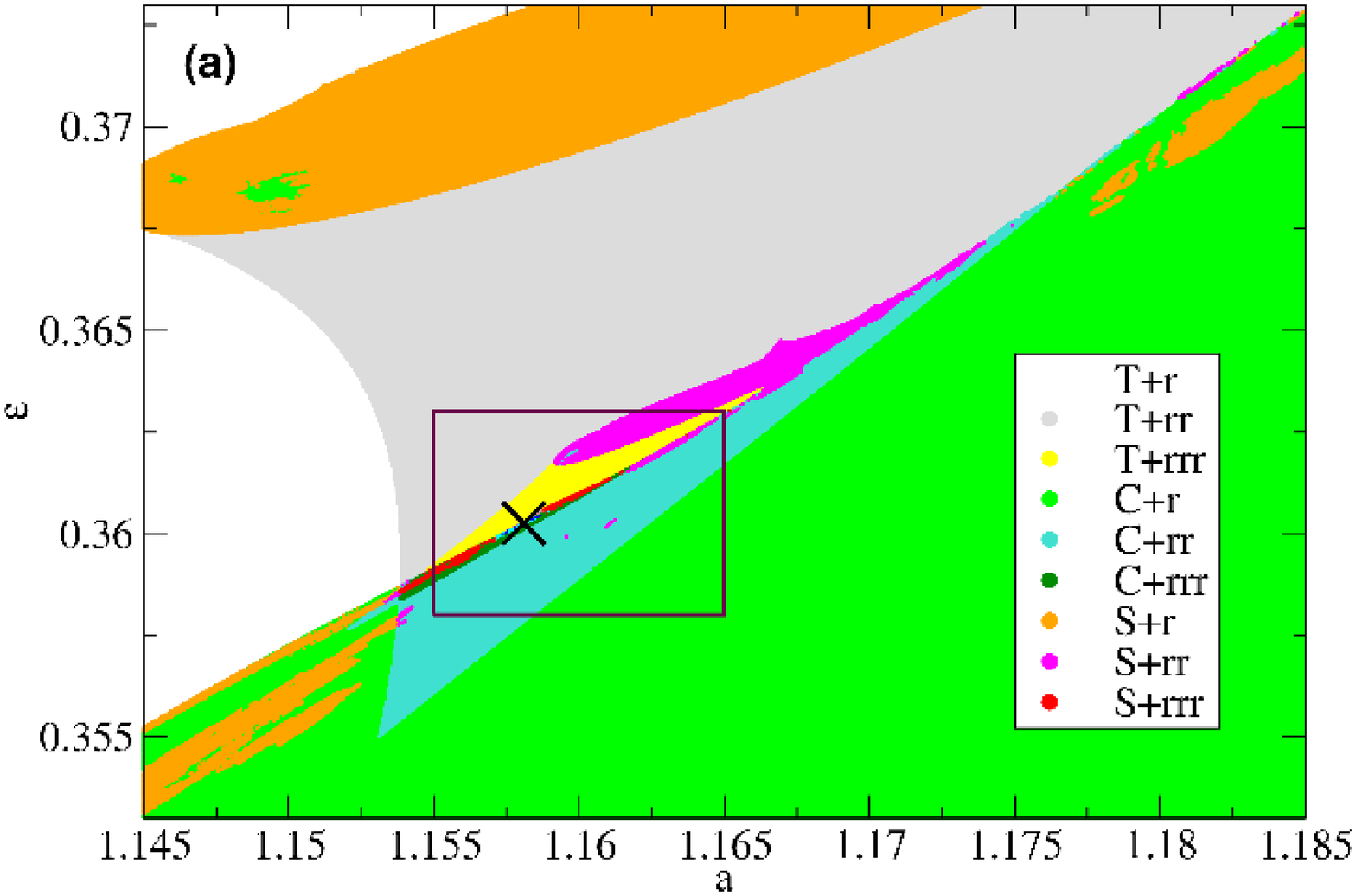}
\includegraphics[clip,width=0.44\textwidth]{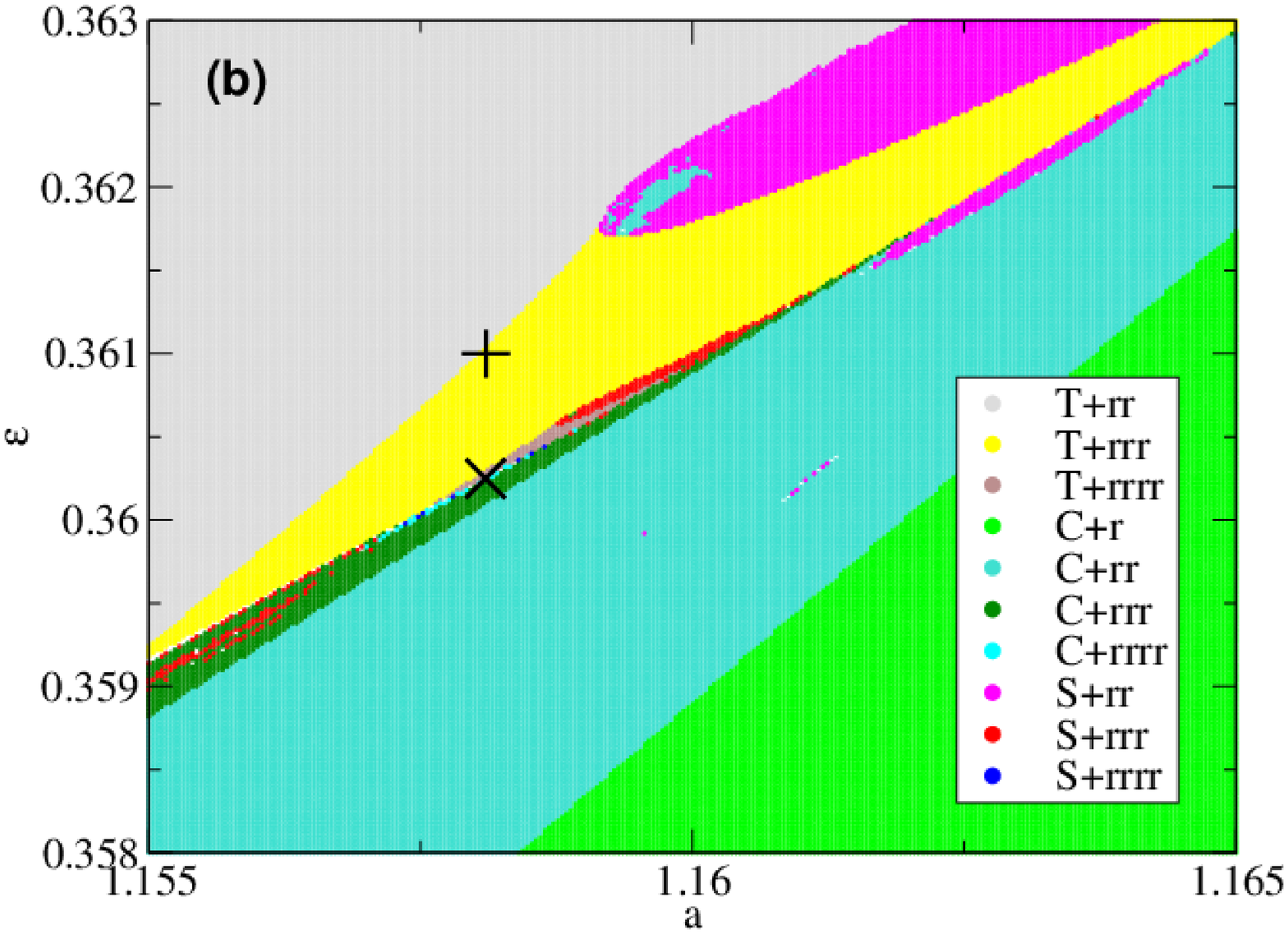}
\caption{Phase diagram near TDT critical point\label{FIG:PD-TDT}}
Black $\times$ indicates the TDT critical point $(1.15809685\dots,
0.36024802\dots)$ (the value obtained in Kuznetsov
etal.\cite{Kuznetsov-ETAL-98}). Black + in box (b) indicates the
parameter value used in the next figure.  ``+rr'', ``+rrr'' etc. in
legends indicate the existence of multiple repellers.  These plots are
obtained with mesh size $(\Delta
a,\Delta\epsilon)$=$(0.00004,0.00002)$.
\end{figure}

\begin{figure}
\includegraphics[clip,width=0.48\textwidth]{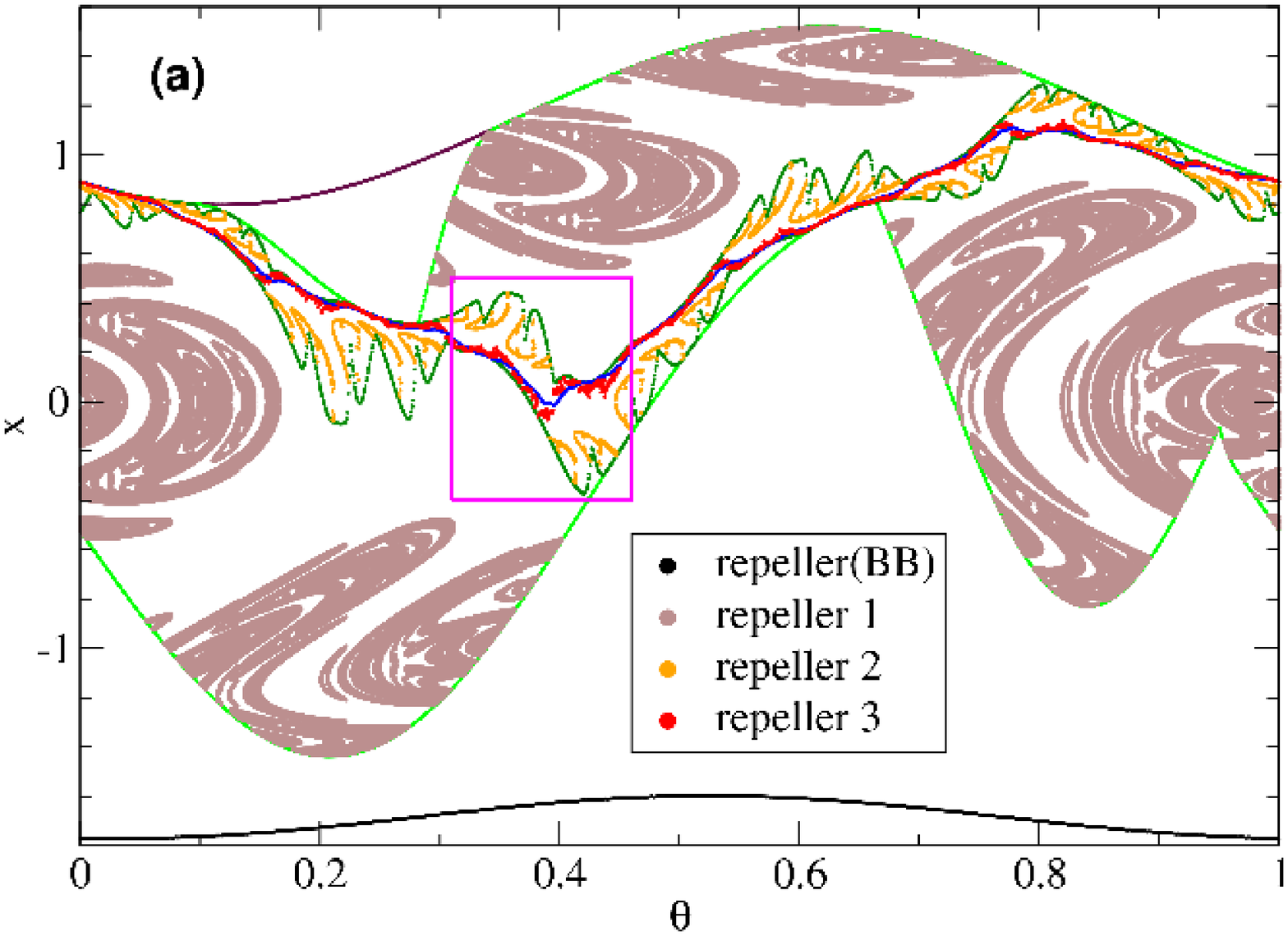}
\includegraphics[clip,width=0.48\textwidth]{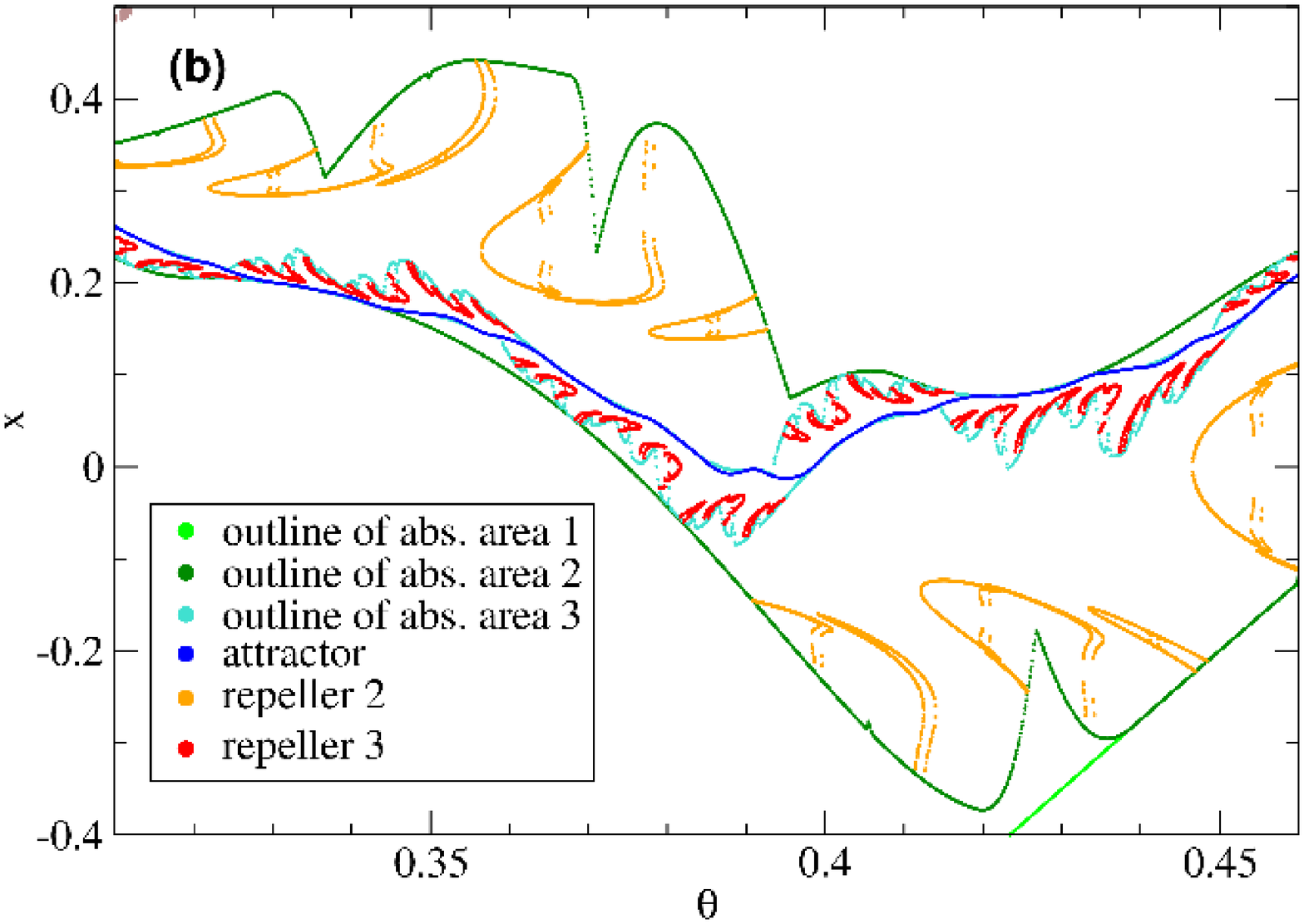}
\caption{Nested structure of absorbing areas (basin of repulsion) and
  corresponding repellers observed at $(a,\epsilon)=(1.1581,0.361)$
  (indicated by the black + in fig. \ref{FIG:PD-TDT}) Plots in box
  (a), (b) are obtained with 2000, 1500 target fibers of $\theta$
  values within $[0,1]$, $[0.31,0.46]$
  respectively. \label{FIG:multi-repellers}}
\end{figure}

An interesting repetitive bifurcation structure has been reported near
the torus doubling termination (TDT) critical
point\cite{Kuznetsov-ETAL-98,Kuznetsov-05}.  Near this critical point,
multiply nested structure of invariant sets is observed as shown in
Figs.\ref{FIG:PD-TDT} and \ref{FIG:multi-repellers}.  The obtained
result is consistent with coexistence of multiple noisy attractors
with different scales\cite{Kuznetsov-05} and also with the features of
the parameter dependence of the attractor\cite{Feudel-etal-06}, i.e.,
the birth of SNA via internal crisis like behavior with decreasing $a$
and via fractalization like behavior with increasing $a$.

In the next figure (Fig. \ref{FIG:fractalized}), the parameter is set
as $(a,\epsilon)=(0.75,0.46672)$, a little after the onset of SNA by
``fractalization'' on the path corresponding to the one investigated
by Kaneko and Nishikawa\cite{Kaneko-84b,Nishikawa-Kaneko-96}.  We can
see a kind of fractal like complicated structure with some apparent
smoothness, however, the images of the repeller is ``dirty'' due to
apparently superimposed irregular vertical stripe like structure.  We
think that the repeller would be dense in its basin of repulsion due
to these stripes and correspondingly the basin of SNA in $\tilde M$
should be riddled, while structures with some smoothness would be
apparent as far as we observe finite number of fibers.

\begin{figure}
\includegraphics[clip,width=0.46\textwidth]{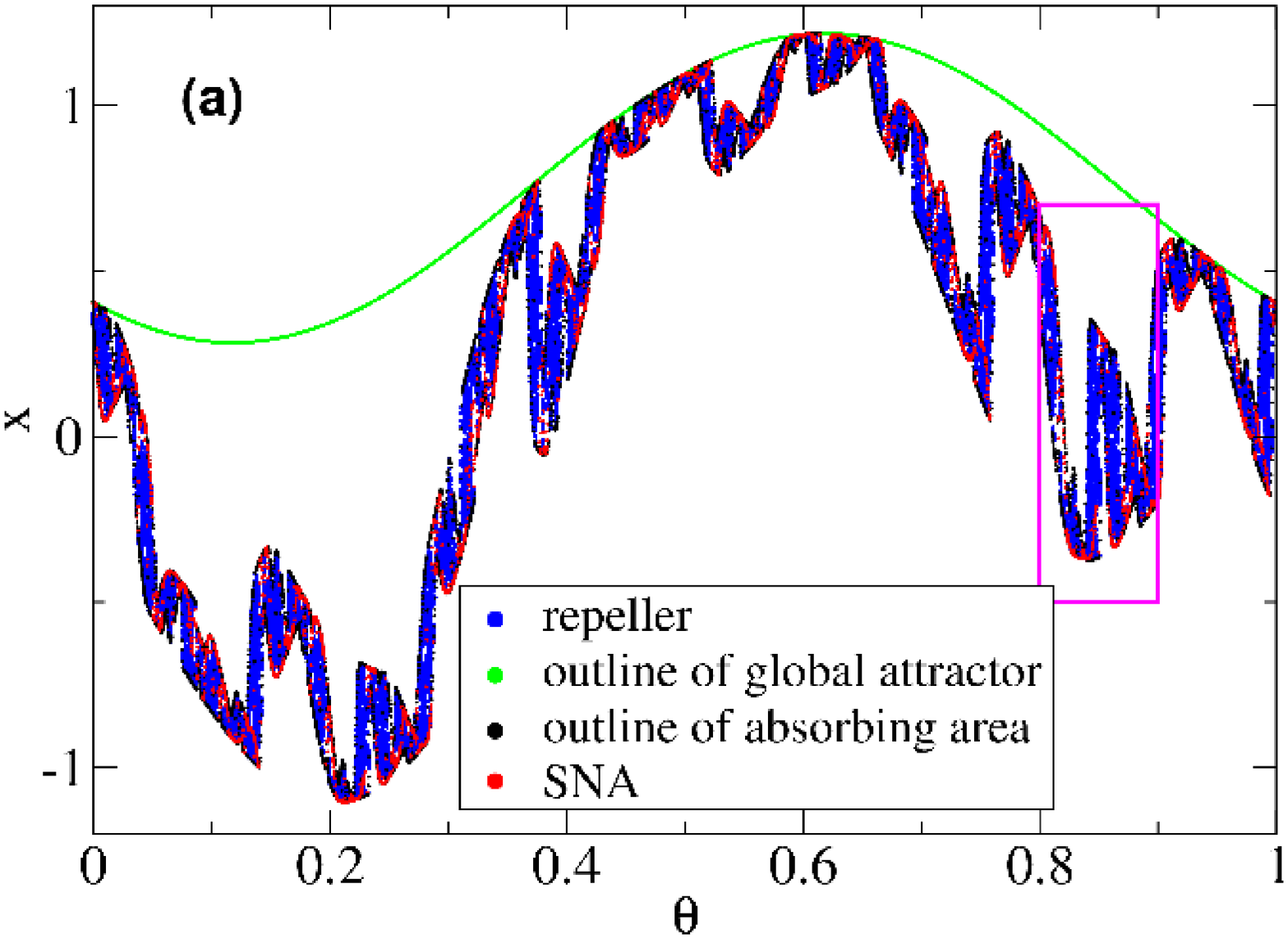}
\includegraphics[clip,width=0.46\textwidth]{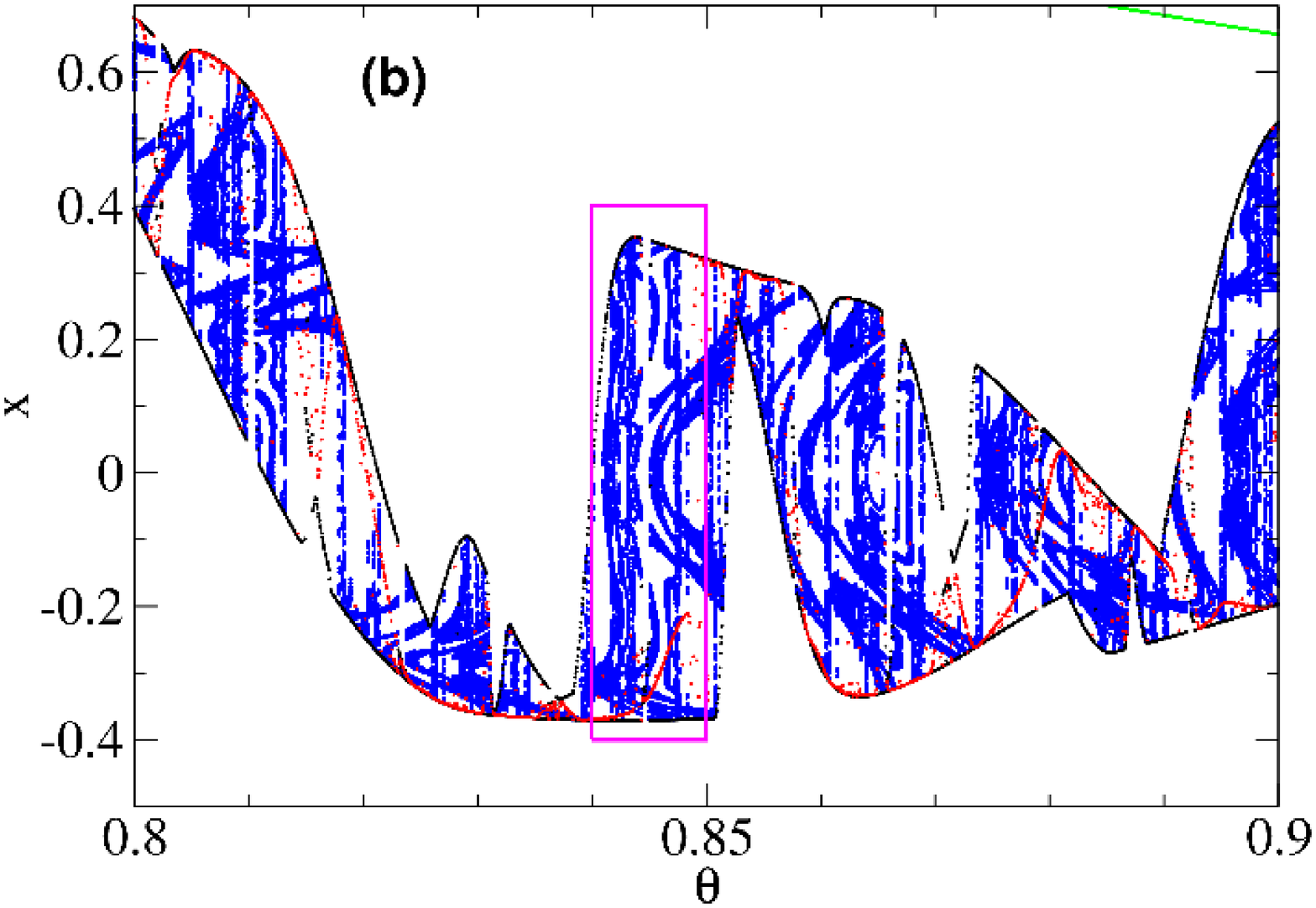}
\includegraphics[clip,width=0.46\textwidth]{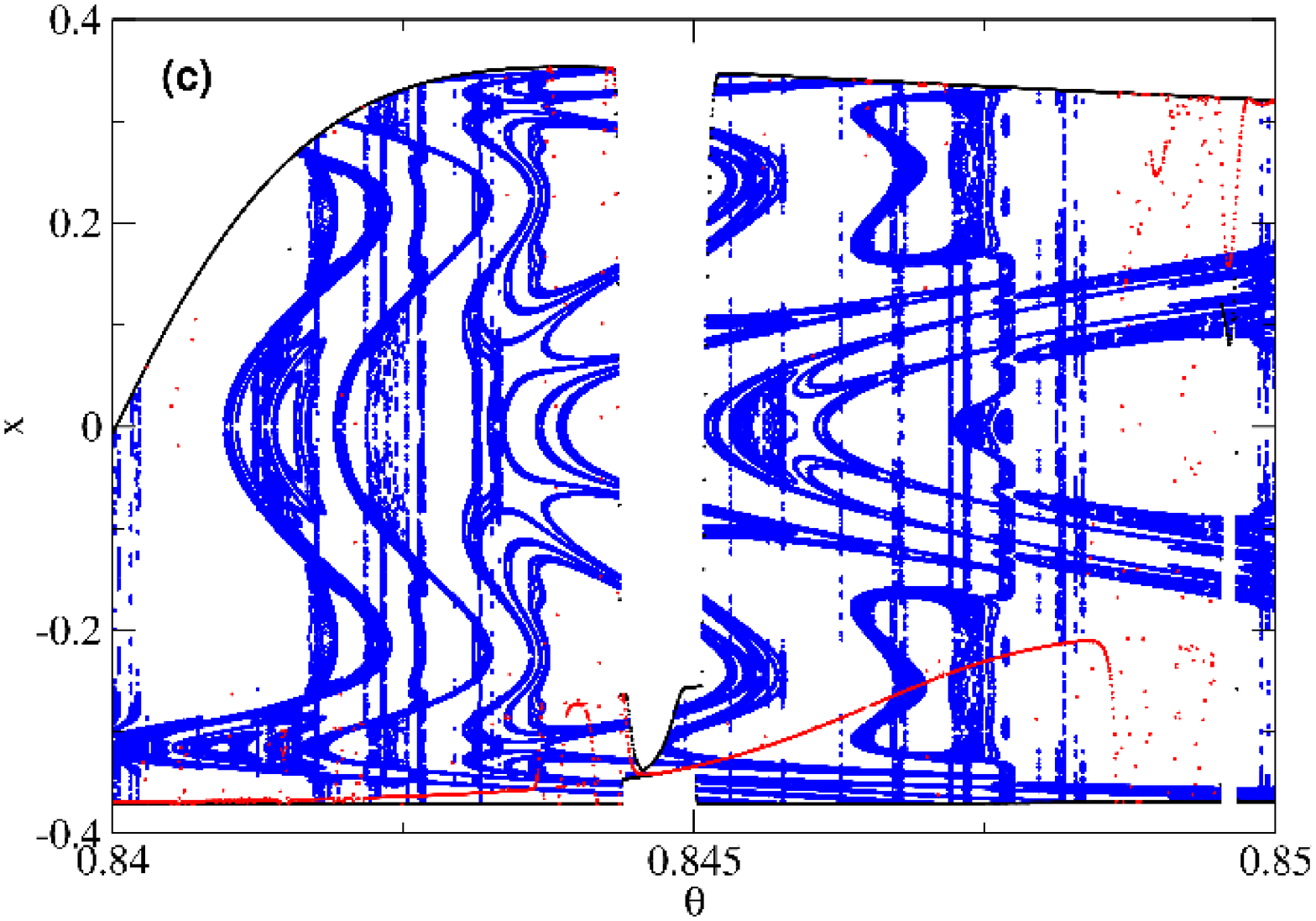}
\caption{Repeller and SNA emerged via fractalization route,
  $(a,\epsilon)=(0.75, 0.46673)$. The pull-back attractor (SNA) and
  the repeller, as well as boundary of absorbing areas, are plotted
  for 4000 (a), 2000 (b and c) target fibers, with $\theta$ values in
  $[0,1]$ for (a), $[0.8,0.9]$ for (b), and $[0.84,0.85]$ for (c)
  respectively.\label{FIG:fractalized}}
\end{figure}

\section{summary}

We proposed a method to observe a kind of ``repeller'' in
quasi-periodically forced logistic map system and a method to
distinguish ``smooth'' attractors and SNAs, using auxiliary
dynamical system describing evolution of a ``line segment'' on a
fiber.  This method enables to obtain images of repellers and
bifurcation diagrams efficiently.

Some numerical results obtained with these methods are exhibited,
namely,
\begin{itemize}
\item phase diagram with some revision in the borderline between
  smooth torus attractor and SNA,
\item detailed images of a repellers with non-trivial ring-like
  structure,
\item absorbing area surrounded by curves with many peaks observed
  nearly at the emergence of a non-trivial repeller, which exhibits a
  symptom of non-uniform saddle-node bifurcation in the segment map
  system,
\item nested structure of multiple absorbing areas and associated
  multiple repellers observed near TDT critical point,
\item image of repeller with apparently noisy structure which is the
  counterpart of SNA.
\end{itemize}

By the use of segment map a similarity between attractors with riddled
basin and SNAs become apparent, and the relevance between various
phenomena including ``fractalization of torus'' and ``emergence of
ring-like shaped repeller'' and (easily observable) bifurcations in
the segment map (i.e., bifurcations accompanied with qualitative
changes of attractors) are also indicated.

At parameter values where the ``fractalization'' occur, it seems that
the transformation from a smooth torus to SNA and the emergence of a
new repeller occur simultaneously.  We will report detailed analysis
on this phenomenon in the forthcoming paper.

Our method make the most of the fact that the dynamics on the fiber is
one dimensional map.  Thus it would not be straight forward to apply
our method to systems with higher dimensional fiber dynamics. On the
other hand, the dynamics of the forcing could be easily generalized,
for example, to chaotic ones.  As the attraction to SNA in
quasi-periodically forced systems could be regarded as a special class
of the weak generalized synchronization phenomena
\cite{Singh-etal-08,Keller-Otani-13,Keller-etal-13}, application of our method to
systems with chaotic driving would help to obtain some new perspective
for generalized synchronization phenomena between non-identical
chaotic elements.

We appreciate N.~Takahashi, T.~Mitsui, and Y.~Sato for helpful
suggestion and discussion.  This research is supported by JSPS KAKENHI
Grant Number 60115938.

\appendix
\section{numerical algorithm}

Here we describe some details of the algorithm used in our numerical
calculation.  Our calculation consists of 3 parts. (1) Search for the
invariant sets with nested structure, which gives information of the
existence/absence of repellers. (2) Decomposition of the invariant
sets into the repeller and the basin of smaller attracting invariant
sets, to obtain the approximate images of the repellers. (3) Check for
the behavior of perturbed orbits starting from a point on the
attractors obtained in (1), in order to distinguish ``smooth''
attracting torus and SNA. Both (2) and (3) need the result of (1), but
(2) and (3) could be carried out mutually independently.

In part (1), we fix a target fiber and will make a list of pull back
attractors of $\tilde M$ on the fiber.  Here we intend to list up all
the attractors that may attract some of the sufficiently short
vertical segments.  The obtained list would contain those
corresponding to the basin of repulsion of repellers as well as those
directly corresponding to the attractors of $M$. Each of the obtained
pull-back attractors is represented by a segment on the fiber (whose
length might be zero).  In this step, we use evolution towards the
target fiber from sufficiently long ago.

In part (2), we choose multiple target fibers that can be reached by
forward iteration from the fiber chosen in (1). For each target fiber,
we locate the images of the segments obtained in (1) and then try to
decompose them to obtain the images of the repellers on the fiber.  In
the decomposition, we use forward iteration from the target fiber.

In part (3), we choose the zero-length segments obtained in (1) that
represent pull-back attractors with negative lyapunov exponent, and
calculate forward trajectory from each of the segments using segment
map with small perturbation to $w$ component.  We also calculate the
forward iteration stating from segments that correspond to the basin
of repellers that possibly has contact with the attractor in question.
If these trajectories coincide after sufficiently long transient
steps, that indicates the existence of a fiber on which the distance
between the attractor and the repeller is smaller than the amplitude
of the perturbation, which could be regarded as a signature of the
absence of the asymptotic stability of the attractor, implying that
the attractor would presumably be an ``SNA''.

In the following, we try to illustrate outline of the algorithm in the
form of a pseudo program.  

\begin{verbatim}

global variables (for parameters): 
  Iinit: a segment on the initial fiber
    chosen so that any invariant set has
    non empty intersection with this segment)
  TList: set of integer
    used to specify theta values of fibers to
    be observed

global variables (for results):
  InvList: set of segments
    list of pull-back attractor of the
    segment map on the target fiber of Part1
  Unity[J] (J: a member of InvList): boolean
    true if the invariant set represented
    by J do not have proper subset which is
    an attracting invariant set
  RepellerImage[J,t]
    (J: a member of InvList,
     t: a member of TList): set of segments 
    image of a section of repeller
    whose basin of repulsion corresponds to J
  SNA[J] (J : a member of InvList):  boolean
    true if J is classified as SNA

working variables:
  I*,J*,K*: segments on a fiber
    I*:  on initial fiber
    J*:  on target fiber
    K*:  on future fiber
\end{verbatim}

\paragraph{Part 1: Search for the attracting sets of $\tilde M$}
\begin{verbatim}
Part1Main
 clear InvList
 call P1CheckImage(Iinit)

Procedure P1CheckImage(I)
  J=IntialTransient(I)
  if (J is not in InvList)
    then
      append J in InvList
      if length(J)>0
        then 
          Unity[J]=P1UnityCheck(I,J)
        else 
          Unity[J]=true
  if ( (Unity[J] is false)
      and (length(I) > param_cutoff) )
    then 
      divide I into (I_1, I_2)
      call P1CheckImage(I_1)
      call P1CheckImage(I_2)

boolean P1UnityCheck(I,J) 
  "Take an arbitrary point v in segment I"
  if ("LyapunovExponent(v)"<0)
    then 
      return false
  "Take a segment V, as
     theta(V)=theta(J),
     center(V)=InitialTransient(v),
     length(V)=very short"
  if (IterateMany(V) and
      IterateMany(J) coincide)
    then
      return true
    else
      return false
\end{verbatim}

In this part, we try to obtain a list of pull back attractors of
$\tilde M$ on a target fiber.  We are interested in such attractors
which are also attractors of $M$ (i.e., stable torus/tori, SNA,
chaotic attractor) or which attract orbits with initial conditions
that correspond to short segments representing sufficiently small
neighborhoods of a point on repellers of $M$.

Let the target fiber be specified as $\{\theta=\theta_0\}$
($\theta_0\in[0,1)$), $C_-$ be a constant satisfying $a-C_-^2
+\epsilon < C_-$, and $\tau^-$ be a sufficiently large integer
parameter that gives the duration of initial transient.  Then we take
``initial segment'' (a point in $\tilde\Omega$)
$I_{init}=(\theta_-,0,|C_-|)$ on the initial fiber specified with
$\theta_-=(\theta_0 - \tau^- \omega)\ \rm{mod}\ 1$.  With this choice
of $I_{init}$, every invariant set of $M$ should have non-empty
intersection with $\underline{I_{init}}$.

We calculate the trajectory starting from a segment $I$ (which is set
as $I_{init}$ for the first trial) on the initial fiber to obtain its
image ($J$) on the target fiber, then record $J$ as a member of the
list of pull back attractor on the target fiber (``InvList'').  If it
has non-zero length and does not correspond to ``transitive'' chaotic
attractor, it is presumed that there exist pull back attractor(s)
which corresponds to a proper subset of $J$.  In such case, we divide
the initial segment $I$ into two, and recursively repeat this
calculation with using each fragment of $I$ as initial segment, until
the initial segment becomes sufficiently short.

\paragraph{Part 2: Visualization of repellers}
\begin{verbatim}
Part2Main
  foreach J in InvList
    if (Unity[J] is false)
      then 
        foreach t in TList
          clear tmpRepellerImage
          JT="Iterate_t_times(J)" 
          call P2Decomposition(JT)
          RepellerImage[J,t]=tmpRepellerImage


Procedure P2Decomposition(J)
  K=IterateManytimes(J)
  call P2DecompSub(J,K)

Procedure P2DecompSub(JX,K)
  L=IterateManytimes(JX)
  if (L matches K)
    then
      if (length(JX)<param_cutoff)
      then
        "append JX to tmpRepellerImage"
      else
        "divide JX into (JX1,JX2)"
        call P2DecompSub(JX1,K)
        call P2DecompSub(JX2,K)
\end{verbatim}

In this part, we choose multiple target fibers, and try to obtain the
image of each repeller on each target fiber.  The target fibers are
chosen to be located on the future of the target fiber of part 1.

We firstly obtain a segments ($J$) on the target fiber that
corresponds to the basin of a repeller, and also its image in a
sufficient future($K$).  Then divide the segment on the target fiber
and check if the image of the fragment in the future is expanded up to
coincide with the image of the original segment.  If their images in
the future matches, it implies that the fragment possibly contains
point(s) of the repeller, and in such case we recursively divide the
fragments again and check their future image.  After a sufficient
number of recursive dividing of the segment, we would obtain an
approximate image of the repeller on the fiber, i.e., a set of
sufficiently short fragments that covers the intersection of the
repeller and the target fiber.

\paragraph{Part 3: Sensitivity check for discriminating SNA and torus}
\begin{verbatim}
Part3Main
  foreach J in InvList
    if (length(J) > 0)
      then
        SNA[J]=false
        K[J]=IterateManytimes(J)
  foreach J in InvList
    if (length(J) is 0)
      then
        L[J]=PerturbedIterateManytimes(J)
        if (L[J] matches one of K[*])
          then 
            SNA[J]=true
          else
            SNA[J]=false

\end{verbatim}

In this part, we try to discriminate SNA from smooth torus by checking
the sensitivity of final state against a perturbation in a small
neighborhood of the attractor.

We calculate a sufficiently long trajectory of perturbed map $\tilde
M^+$ starting from a segment (which is the member of InvList in
question).  If the segment corresponds to an asymptotically stable
attractor (i.e., smooth torus/tori attractor), the perturbation would
not kick the trajectory out of the basin of the original attractor,
but if it corresponds to an SNA, the trajectory would be attracted to
another attractor corresponding to the basin of a repeller after a
sufficiently long iteration.

Note that the perturbation $\tilde W^+$ affects the map only in a
neighborhood of $\{w=0\}$ plane, thus attractors detached from
$\{w=0\}$ are not affected by the perturbation.

\end{document}